\documentclass{article}

\usepackage{arxiv}
\usepackage{nicefrac}
\usepackage{url}
\usepackage{textpos}
\usepackage[utf8]{inputenc} % allow utf-8 input
\usepackage[T1]{fontenc}    % use 8-bit T1 fonts
\usepackage{hyperref}       % hyperlinks
\usepackage{url}            % simple URL typesetting
\usepackage{booktabs}       % professional-quality tables
\usepackage{amsfonts}       % blackboard math symbols
\usepackage{nicefrac}       % compact symbols for 1/2, etc.
\usepackage{microtype}      % microtypography
\usepackage{lipsum}         % Can be removed after putting your text content
\usepackage{doi}
\usepackage{graphicx}
\usepackage{float}
\usepackage{subfigure}
\usepackage{epstopdf}
\usepackage[dvips]{epsfig}
\usepackage{lineno,hyperref}
\usepackage{anysize}
\marginsize{3cm}{3cm}{2cm}{2cm}
\usepackage{amscd}
\usepackage{mathrsfs}
\usepackage{verbatim}
\usepackage{graphicx}
\usepackage{amsmath}
\usepackage{cleveref}       % smart cross-referencing
\usepackage{amssymb}
\usepackage{array}
\usepackage{booktabs}
\usepackage{soul,xcolor}

\usepackage{soul}  %%% The command for highlight is \hl{.......}
\usepackage{color}
\usepackage{lineno}
\setstcolor{red}

\usepackage{adjustbox}
\usepackage{mdframed}
\usepackage{multicol}
\usepackage{multirow}
\usepackage{algpseudocode}
\usepackage{cite}

\title{GroupMixer: Patch-based Group Convolutional Neural Network for Breast Cancer Detection from Histopathological Images}

\newif\ifuniqueAffiliation
\uniqueAffiliationtrue

\ifuniqueAffiliation % Standard variant of author block
\author{Ardavan Modarres \\
	Department of Electrical Engineering\\
	K.N. Toosi University of Technology\\
	Tehran, Iran \\
	\texttt{ardavan.modarres@email.kntu.ac.ir} \\
	\And
	Erfan Ebrahim Esfahani \\
	Department of Electrical Engineering\\
	K.N. Toosi University of Technology\\
	Tehran, Iran \\
	\texttt{erfan.esfahani@email.kntu.ac.ir} \\
	\And
	Mahsa Bahrami \\
	Department of Electrical Engineering\\
	K.N. Toosi University of Technology\\
	Tehran, Iran \\
	\texttt{mahsa.bahrami@email.kntu.ac.ir} \\
}

\renewcommand{\shorttitle}{GroupMixer: Patch-based Group Convolutional Neural Network for Breast Cancer Detection} 

\hypersetup{
	pdftitle={GroupMixer: Patch-based Group Convolutional Neural Network for Breast Cancer Detection from Histopathological Images},
	pdfsubject={Breast cancer, computer-aided diagnosis, histopathological images, deep learning, convolutional neural networks, machine learning},
	pdfauthor={Ardavan Modarres, Erfan Ebrahim Esfahani, Mahsa Bahrami},
	pdfkeywords={Breast cancer, computer-aided diagnosis, histopathological images, deep learning, convolutional neural networks},
}

\begin{document}
	\maketitle
	
\begin{abstract}
Diagnosis of breast cancer malignancy at the early stages is a crucial step for controlling its side effects. Histopathological analysis provides a unique opportunity for malignant breast cancer detection. However, such a task would be tedious and time-consuming for the histopathologists. Deep Neural Networks enable us to learn informative features directly from raw histopathological images without manual feature extraction. Although Convolutional Neural Networks (CNNs) have been the dominant architectures in the computer vision realm, Transformer-based architectures have shown promising results in different computer vision tasks. Although harnessing the capability of Transformer-based architectures for medical image analysis seems interesting, these architectures are large, have a significant number of trainable parameters, and require large datasets to be trained on, which are usually rare in the medical domain. It has been claimed and empirically proved that at least part of the superior performance of Transformer-based architectures in Computer Vision domain originates from patch embedding operation. In this paper, we borrowed the previously introduced idea of integrating a fully Convolutional Neural Network architecture with Patch Embedding operation and presented an efficient CNN architecture for breast cancer malignancy detection from histopathological images. Despite the number of parameters that is significantly smaller than other methods, the accuracy performance metrics achieved 97.65\%, 98.92\%, 99.21\%, and 98.01\% for $40x$, $100x$, $200x$, and $400x$ magnifications respectively. We took a step forward and modified the architecture using Group Convolution and Channel Shuffling ideas and reduced the number of trainable parameters even more with a negligible decline in performance and achieved 95.42\%, 98.16\%, 96.05\%, and 97.92\% accuracy for the mentioned magnifications respectively.
	\end{abstract}
	
% keywords can be removed

\section{Introduction and background}
\paragraph{}
Breast cancer (BC) is one of the most common types of cancer diagnosed in women, second only to skin cancer \cite{Mayo,ACS}. Although BC can also occur in men, it is far more frequent in women \cite{who}. From 2005 to 2020, the number of new annual cases of BC has increased from 1.4 million to 2.3 million \cite{who,IRBM} and is hardly projected to decrease by 2025 \cite{IRBM}. According to American Cancer Society statistics, on average, 1 out of 8 women develop BC at some stage of their life \cite{ACS}. With a fatality rate of 2.5\%, BC is the second deadliest incarnation of cancer, after lung cancer \cite{ACS}.
\paragraph{}
The gold standard for early diagnosis and screening for BC is mammography, often but not always complemented by breast \textcolor{black}{ultrasound} or \textcolor{black}{magnetic resonance imaging}  \cite{Mayo,who,ACS}. Recently, research on \textcolor{black}{computed tomography} as another diagnostic tool for BC detection has started to gain momentum \cite{CT1, CT2}. While all these imaging modalities are invaluable tools for screening and non-invasive early diagnosis, they are not always the most effective in cases where BC is already known to exist, and the only question left is whether the tumor is benign or malignant. In such cases, the usual approach, although invasive and unpleasant, is to collect a sample from the diseased tissue using special needles (a procedure known as biopsy) and study the sample under a microscope. Traditionally, the lab analysis of histopathology images has been carried out manually by pathologists who examine the images closely, looking for malignancy patterns. Obviously, such a task is slow and requires \textcolor{black}{expertise}, which can not be immediately available in many settings \cite{ResNet18-TL}. This issue leads to long pending times, causing potential ramifications for the patient.
Realizing such limitations, researchers have shifted focus to \textcolor{black}{Machine Learning} and \textcolor{black}{Computer-Aided Diagnostic} systems to help speed up the diagnosis and reduce human error. 
\paragraph{}
\textcolor{black}{Over the past two decades, conventional machine learning approaches have been attempted for the classification of Histopathological Breast Cancer (HBC) images. The study of \cite{81} focused on classifying 1000 HBC images using the K-Nearest Neighbors (KNN) approach based on morphology features extracted from the images.  In \cite{69}, Support Vector Machines (SVMs) were explored to classify 48 breast tissue images. The approach was based on dimensionality reduction and achieved high performance. However, a major drawback was the small number of images considered in the study. The work of \cite{47} also used SVMs to classify normal, benign, and malignant cell images. While the approach obtained informative features for classification and also reported high performance metrics, this work also considered only 68 HBC images, which made the result less likely to generalize well. The method of \cite{74} used a hybrid strategy that combined SVM, KNN, and Probabilistic Neural Network classifiers on extracted features by applying Principal Component Analysis (PCA). In \cite{79}, salient features characterizing the tumors were extracted from a multitude of fields of view. Afterward, a Random Forest (RF) classifier was utilized to distinguish low, intermediate, and high-grade tumors. While all these conventional approaches (and many others reviewed in \cite{IRBM, JMIRS}) were insightful and inspiring, a common pitfall about them is the costly requirement of domain expert knowledge for the manual extraction of informative features for performing classification tasks effectively.}
\paragraph{}
Deep Learning (DL) \cite{DL} based methods have recently been proposed to classify HBC images. One of the major improvements of DL over traditional approaches is the automatic extraction of discriminative features for classification without requiring domain knowledge. For classifying HBC images, the publicly available BreakHis dataset \cite{breakHis} is one of the largest of its kind and a benchmark for training such DL models. Numerous \textcolor{black}{solutions based on Deep Neural Network architectures} have been trained on this dataset for HBC classification in the recent literature. For instance, in \cite{fusion}, a two-\textcolor{black}{branch} Neural Network comprising a branch with standard Convolutional layers and a branch with Capsule Networks was used to extract and fuse features for HBC image classification. The \textcolor{black}{Convolutional Neural Network (CNN)} \textcolor{black}{branch} contained three repeating modules, each comprising two consecutive Convolutional layers and one layer of Pooling. The Capsule \textcolor{black}{branch} also contained three consecutive Convolutions, followed by a PrimaryCaps unit. The algorithm was efficient and achieved state-of-the-art classification results. \textcolor{black}{A modified version of ResNet-34 and ResNet-50 was developed in \cite{MResNet}.} \textcolor{black}{This approach seemed to be effective and achieved competitive results versus the state-of-the-art methods. However, a limitation of that work was the lack of a comparison with the results the original ResNet architecture without any modifications could achieve. It was inconclusive whether the good performance of the modified ResNets was due to the changes made in the architecture by the authors or just the pre-trained backbone of ResNet itself. The work of \cite{DRDA} was inspired by the bottleneck module of the ShuffleNet architecture \cite{Shuffle v2} and included three major building blocks: a Dual-Shuffle Residual block, a Channel-Attention block, and a Residual Dual-Shuffle Attention block. While the first two blocks of this architecture are different, unique building blocks, the third one is built from the first two. A dense connection of these blocks resolved the overfitting and vanishing gradient problems and gave the overall model significant ability in binary classification of HBC images. A public web application was also developed based on this method. In \cite{Razi}, a pair of Deep Neural Networks, known as the Teacher-Student framework, were developed in which the teacher network was trained on all four magnification factors of the BreakHis dataset while the student was trained only on one of the magnification factors and the knowledge of the remaining three categories was transferred from the teacher to the student within the Knowledge Distillation framework. This approach helped the model remain lightweight yet effective in classifying unseen examples.}
\paragraph{}
Since the BreakHis dataset can not be regarded as large compared to other famous DL datasets such as ImageNet \cite{imagenet}, which includes over a million training images, harnessing the power of pre-trained networks trained on large datasets and Transfer Learning (TL) is also among popular approaches within the community. In \cite{91}, a pre-trained GoogLeNet \cite{GoogleNet} model, originally trained on ImageNet, was adopted, and the last layers of the architecture were replaced by a convolutional layer followed by pooling and a classification head. In order to further compensate for the lack of sufficient HBC data, data augmentation was also employed in this work. However, based on the report in \cite{91}, it seems that the augmentation has taken place over all data rather than only the training data, which causes data leakage issues and makes the results less reliable. In \cite{ResNet18-TL}, a pre-trained ResNet \cite{original_resnet}, originally trained on the ImageNet dataset, was utilized and fine-tuned on the BreakHis dataset. In this regard, a couple of fully connected and two Convolutional layers were trained while the others were frozen. Furthermore, this method exploited valid data augmentation by flipping and rotating the training images. The research of \cite{Rahhal} was another TL based approach, wherein five successive Convolutional layers of a pre-trained architecture similar to VGG \cite{VGG} network, which was initially trained on ImageNet, served as feature a extractor for a classifier with two fully connected layers. The classifier weights were fine-tuned on the BreakHis dataset. In \cite{histok}, a generalizable knowledge between histopathological datasets was sought, a data-centric transfer learning approach was utilized, and weight distillation was used to share knowledge between models without the additional cost of carrying out training procedure. More comprehensive reviews on recent DL-based classification algorithms for BC detection from histopathological images are conducted in \cite{IRBM, JMIRS}. Moreover, an exhaustive survey on recent developments in deep learning and computer vision algorithms for classifying medical images and Computational Pathology is carried out in \cite{copath}.
\paragraph{}
All the works discussed above fall into the category of CNNs. CNNs have been the dominant architecture for image analysis and vision tasks in recent years. However, recently it has been speculated that the Transformer-based architectures \cite{transformer} such as Vision Transformer (ViT) \cite{ViT}, borrowing the Attention Mechanism \cite{attmech} which has already shown impressive success in \textcolor{black}{Natural Language Processing (NLP)}, may become the new image analysis revolution and dominate CNNs as well, at least in some specific tasks and domains. Indeed, Transformer-based architectures have shown greater performance than some famous CNNs such as ResNet \cite{ViT} in some design choices and under certain conditions. The main building block of Transformer-based architectures is the Self-Attention Mechanism, which suffers from quadratic computational complexity. To alleviate this issue and make the Self-Attention Mechanism more cost-effective, Transformer-based architectures in computer vision tasks such as classification utilize patch embedding operation at first, which is dividing the input image into small fixed-size non-overlapping regions and linearly projecting these patches into a lower-dimensional space. Beyond the issue of Self-Attention computational complexity, Transformer-based architectures have a significant number of trainable parameters, and training a Transformer-based architecture requires a large dataset. However, a motivating question is whether the superior performance of Transformer-based architectures in image analysis originates from the more powerful architecture of the Transformer and Self-Attention Mechanism, or at least to some degree owned to the patch embedding operation. An attempt has been made in \cite{Patches} to partially answer this question by integrating the Patch Embedding operation into a fully Convolutional Neural architecture for image classification tasks. This strategy, which leverages the merits of Transforme-based architectures, was evaluated on ImageNet benchmark and outperformed some of state-of-the-art Convolutional and even Transformer-based architectures in image analysis realm such as ResNet, DeiT \cite{deit} and ResMLP \cite{resmlp} with significantly more parameters, suggesting that the excellent performance of Transformer-based architectures (which was also equipped with patch embedding), at least to some degree, comes from the patch embedding operation.
\paragraph{}
\textbf{Conributions}: We were motivated to harness the power of Transformer-based architectures and leverage the merits of these powerful architectures.
Due to the lack of existence of enormous datasets for HBC, we were motivated to:
\begin{itemize}
	\item Borrow the idea of equipping a fully Convolutional Neural Network with patch embedding operation in this research, and develop and present a light and efficient Convolutional architecture based on the abstract architecture proposed in \cite{Patches}.
	\item Moreover, we took a step forward and proposed to modify the abstract architecture proposed in \cite{Patches} by utilizing Group Convolution and Channel Shuffling ideas proposed in \cite{shufflev1}, and we reduced the number of trainable parameters even more with a slight decline in performance.
\end{itemize}
The rest of this paper is organized as follows: Section \ref{sec proposed framework} introduces the proposed framework for BC detection from histopathological images, and the utilized dataset. Section \ref{Exp} reports the experiments and results, and Section \ref{sec concl} draws the conclusions.
\section{Material and Methods}\label{sec proposed framework}
\subsection{Dataset}
\begin{figure*}[!t]
\includegraphics[width=\textwidth]{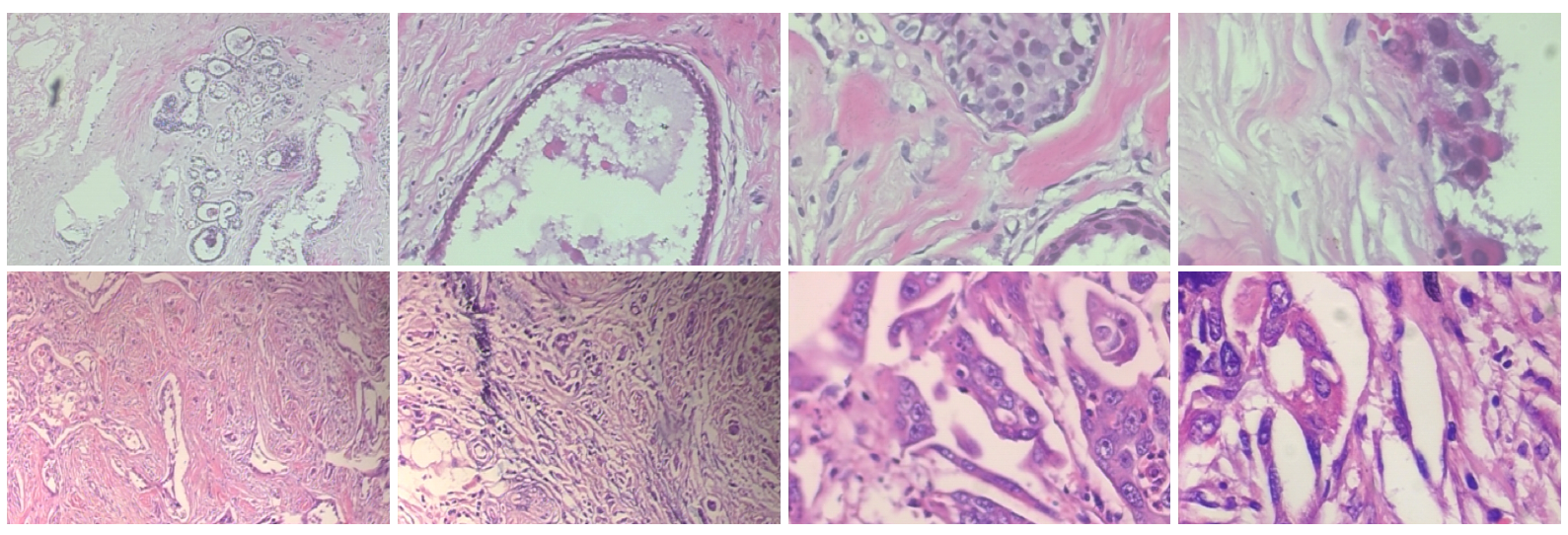} \caption{Instances of breast cancer histopathological images. top row: breasts with benign state in ($40\text{x}$, $100\text{x}$, $200\text{x}$, $400\text{x}$ magnifications respectively); buttom row: breasts with malignant
state in ($40\text{x}$, $100\text{x}$, $200\text{x}$, $400\text{x}$ magnifications respectively).}
\label{Samples}
\end{figure*}

\paragraph{}
In this paper, the BreakHis \cite{breakHis} dataset, which was collected in the P\&D Laboratory of Brazil, was used to train and evaluate the presented method. BreakHis dataset contains 7909 HBC images from histopathological samples of Benign and Malignant stages of breast cancer in four resolutions, namely, $40\text{x}$, $100\text{x}$, $200\text{x}$, and $400\text{x}$ collected from 82 patients. Each of the Benign and Malignant classes contain 5429 and 2480 images, respectively.
\pagebreak \\
There seems to be a slight imbalancement between the two classes, and it is an important issue to develop a classifier without any bias on the class with more samples. Figure \ref{Samples} illustrates one example of a benign sample and one example of a malignant sample for each resolution. The details for this dataset can be found in \cite{breakHis}. 
\subsection{Presented Method}
\paragraph{}
Due to lots of achievements of Transformer-based architecture in the Computer Vision domain, we were highly motivated to leverage the benefits of these models in Medical Image Analysis, as previously mentioned, and improve the performance of Diagnosis Systems. Due to the lack of existence of large medical datasets, especially for HBC classification, we adopted the idea of equipping a Convolutional architecture with Patch Embedding operation and presented a light and efficient CNN architecture based on the abstract architecture proposed in \cite{Patches} as our presented base model. Beyond that, we borrowed Group Convolution and Channel Shuffling ideas proposed in \cite{shufflev1}, and we proposed our slim models by modifying the base model, which achieved excellent results and outperformed some of the research in the literature with much less number of parameters.
\subsubsection{Base Model}
\paragraph{}
The presented base model for classifying HBC images starts with a patch embedding layer at the beginning. The $c$-channel input HBC images are divided into $h$ small patches of size $p$. Then, these patches are nonlinearly projected to a lower-dimensional space by applying a linear transformation followed by a nonlinear activation function such as ReLU \cite{relu} and Batch Normalization (BN) \cite{batchnorm} layer. The described patch embedding operation can be implemented by a Convolution layer with $c$ channels (embedding dimension $c$) with stride $p$ and kernel size $p$ as described in Equation \ref{eq1}.
\begin{equation}\label{eq1}
f_0 = \text{BN}(\sigma(\text{Conv}_{c\rightarrow h}(X, \text{Stride}=p,\text{Kernel Size}=p)))
\end{equation}
Where $f_0$ is the embedding of the patches and is input to the rest of the network and $\sigma$ is a nonlinear activation function, ReLU is used in current research. The rest of the network is several successive application of ConvMixer layer, which comprises a Depthwise Convolution along the depth dimension (to extract spatial features) followed by subsequent nonlinear activation function and BN, skip connection bypassing the input feature map of Depthwise Convolution to output of it, and a Pointwise Convolution (i.e., kernel size $1\times 1$) to combine extracted features by Depthwise Convolution, again followed by nonlinear activation function and BN as described in Equations \ref{eq2}, \ref{eq3}.
\begin{equation}\label{eq2}
	f'_l = \text{BN}(\sigma(\text{DepthwiseConv}(f_{l-1}))) + f_{l-1}
\end{equation}
\begin{equation}\label{eq3}
	f_{l} = \text{BN}(\sigma(\text{PonitwiseConv}(f'_l)))	
\end{equation}
Where $f_{l-1}$ is the output of Pointwise Convolution in the previous ConvMixer layer, $f'_l$ is the input of the Pointwise Convolution of current ConvMixer layer, $f_{l}$ is the output of the Pointwise Convolution of current ConvMixer layer, and $\sigma$ is ReLU activation function.
\begin{figure*}[t!]
	\includegraphics[width=\textwidth]{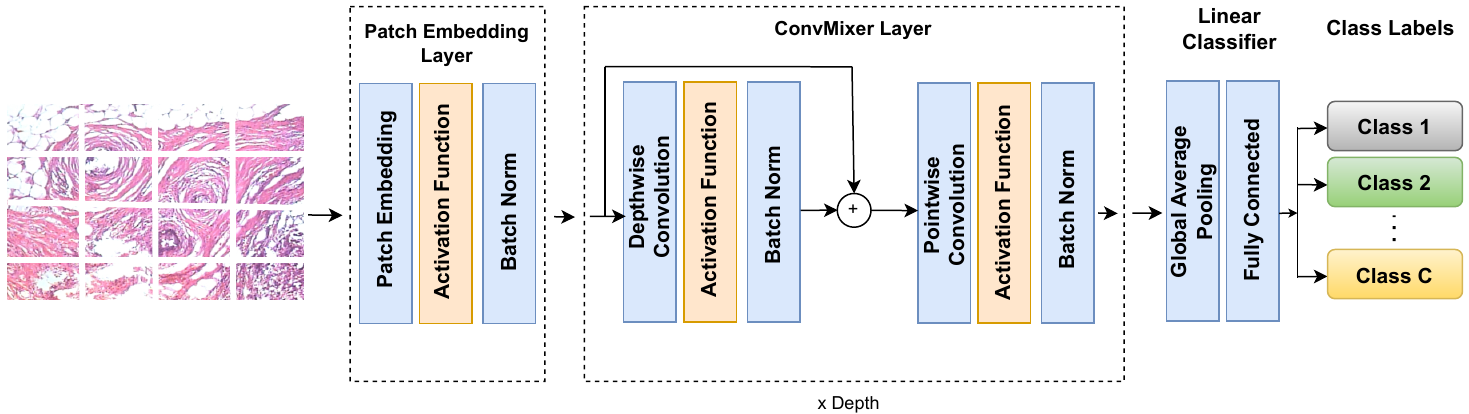} \caption{The presented architecture \cite{Patches} for classifying HBC images.}\label{Model}
\end{figure*}
\paragraph{}
The incentive behind this choice of Convolutional layers is having a large receptive by applying Depthwise Separable Convolutions, which are much more efficient by factorizing standard Convolution operation in two feature extraction and feature fusion phases through Depthwise and Pointwise Convolution operations, respectively, using much fewer parameters than standard Convolution. Upon multiple applications of the ConvMixer ($\times d$ times), the output is passed to a Global Average Pooling layer to shrink down to a feature vector of size $h$ and finally given to a simple linear classifier comprised of only one fully connected layer. The architecture is depicted in Figure \ref{Model}. In the next step, the Group Convolution and Channel Shuffling operations are utilized to decrease the number of trainable parameters even more and design slim architectures. Channel Shuffling enables information exchange across different groups in order to enhance network representation capability. It shuffles feature maps along the channel dimension, allowing each group to access the information of all other groups. This operation helps improve network performance without introducing additional computational costs.
\subsubsection{Slim Model}
\begin{figure}[h!]
	\centering
	\includegraphics[width=0.45\textwidth]{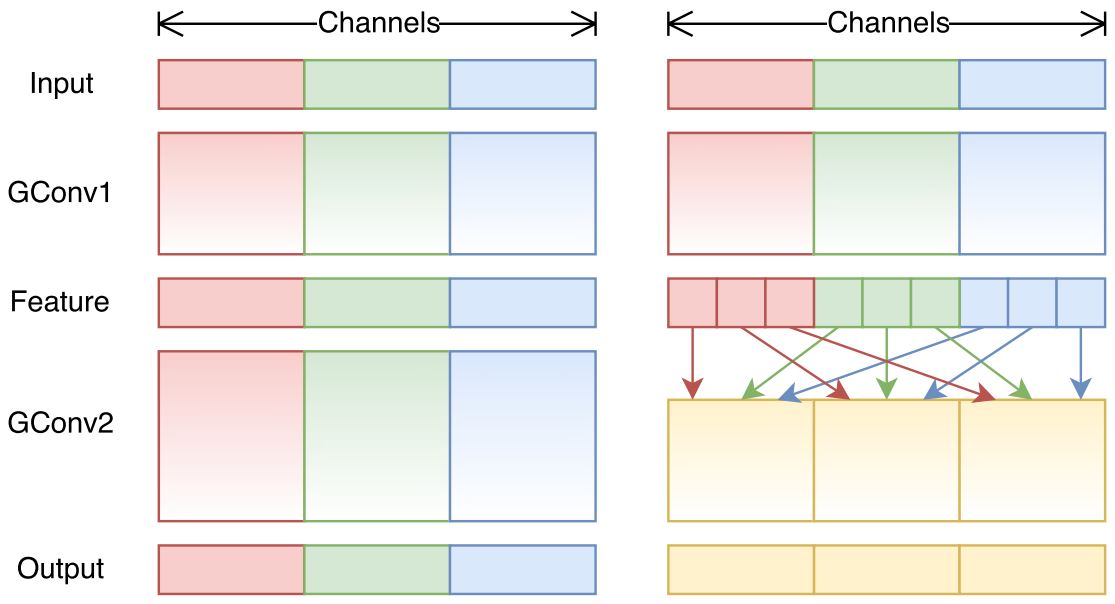}\caption{Utilizing channel shuffle with two consecutive group convolutional layers. GConv denotes group convolution. a) Employing two stacked convolutional layers with an equal number of groups. Each resulting output channel exclusively corresponds to input channels within the same group, no inter-group communication; b) When GConv2 follows GConv1, input and output channels are completely interconnected across different groups. \cite{shufflev1}.}
	\label{channelshuffle}
\end{figure}

\paragraph{}
Since many medical datasets, such as HBC datasets and BreakHis dataset specifically, are not as large as benchmark datasets such as ImageNet, training large neural networks with lots of trainable parameters leads to overfitting, which negatively affects the training process and decreases the upper-bound performance of the utilized model. Moreover, large models with millions of parameters require more expensive computational resources for training and inference procedures, leading to more expensive histopathological analysis for patients. In order to overcome these issues, As previously mentioned, we utilized the idea of Channel Shuffling and Group Convolution in order to achieve a more efficient architecture with as little decline in performance as possible. By applying Group Pointwise Convolution instead of standard Pointstwise Convolution, we can simultaneously reduce the number of parameters and the computational complexity. The Group Convolution and Channel Shuffling schemes are depicted in Figure \ref{channelshuffle}.
	
\begin{figure*}[t!]
	\includegraphics[width=\textwidth]{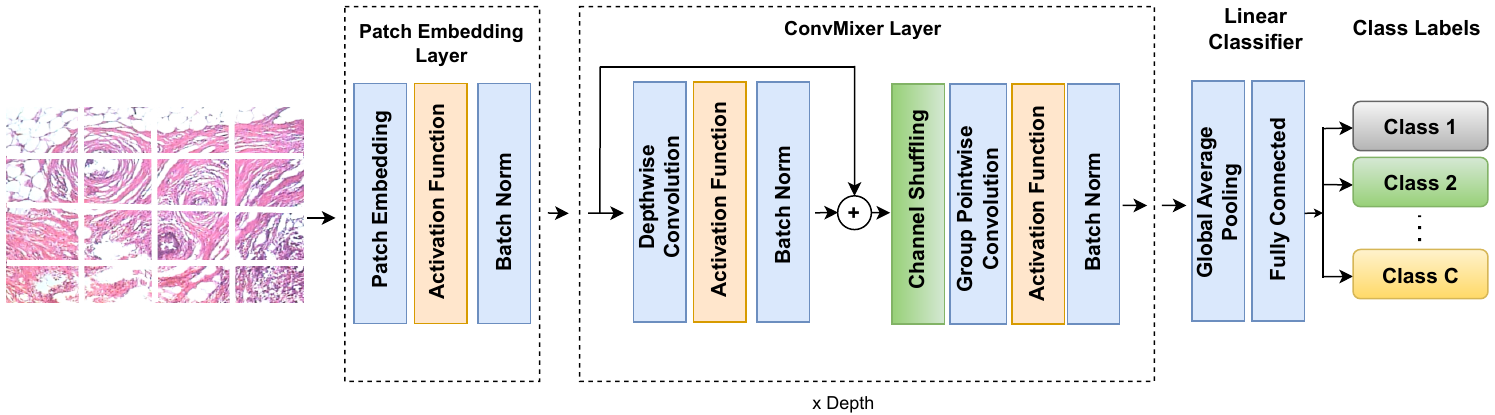}\caption{Slim architecture}
	\label{slimmodel}
\end{figure*}

The architecture integrating Channel Shuffling and Group Convolution operations is depicted in Figure \ref{slimmodel}. \pagebreak \\
We have developed Slim models with two settings:% Group Size $=2$ and Group Size $=4$.
\begin{itemize}
	\item Group Size $=$ 2
	\item Group Size $=$ 4
\end{itemize}
While larger Group Size leads to a lighter model, it increases the risk of information loss. The following section compares the performance of these networks with each other and with state-of-the-art HBC classification methods.
\section{Experiments}\label{Exp}
\subsection{Training setup}
\paragraph{}
The training, validation, and test sets were generated by randomly dividing the whole dataset into 70\%, 20\%, and 10\% splits, respectively. Random horizontal and vertical flips were carried out as data augmentation transformations. The hyperparameters of the proposed method were set as follows: kernel size = $7$, stride = $7$ for $128$ convolution filters within the patch embedding layer, kernel size = $9$, stride = $1$, number of convolution filters = $128$ in each ConvMixer layer. Each of the slim models and the base model contain three ConvMixer layers. We found these sub-optimal hyperparameters by a minimal search over them. The network was trained by minimizing the focal loss function with Adam optimizer with learning rate = $4\times10^{-3}$. We utilized early stopping to train each model for an appropriate number of epochs and prevent overfitting.
\subsection{Results and discussion}
\begin{figure}[h!]
		\centering
		\subfigure[40x]{\includegraphics[width=0.23\textwidth]{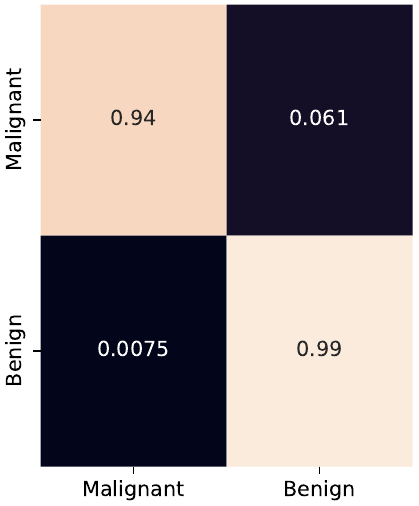}}
		\hfill
		\subfigure[100x]{\includegraphics[width=0.23\textwidth]{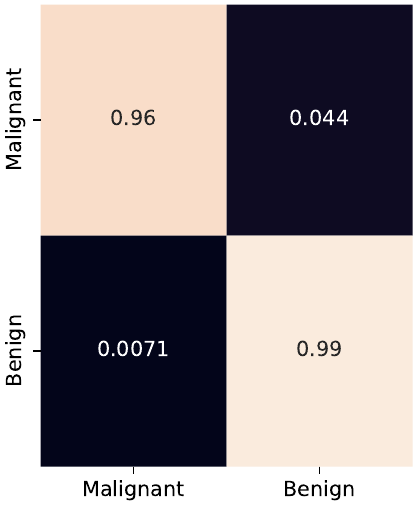}}
		\hfill
		\subfigure[200x]{\includegraphics[width=0.23\textwidth]{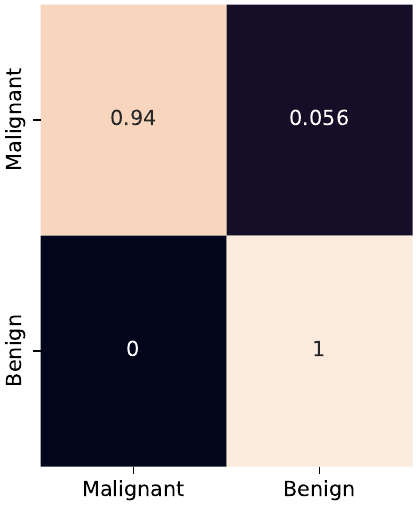}}
		\hfill
		\subfigure[400x]{\includegraphics[width=0.23\textwidth]{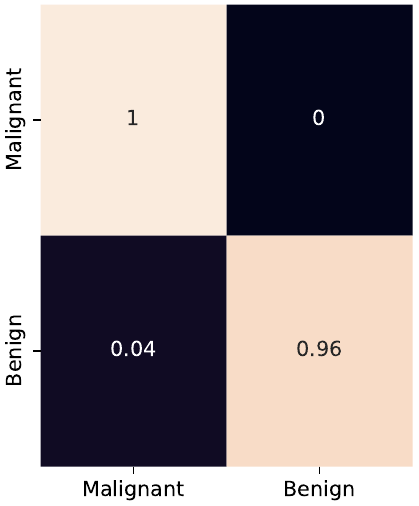}}
		\caption{Confusion matrices for the presented base model for 40x, 100x, 200x, and 400x zoom factors.}\label{BCMs}
	\end{figure}

\begin{figure}[h!]
		\centering
		\subfigure[40x]{\includegraphics[width=0.23\textwidth]{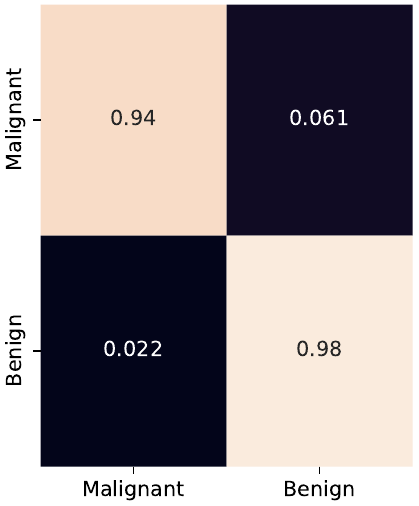}}
		\hfill
		\subfigure[100x]{\includegraphics[width=0.23\textwidth]{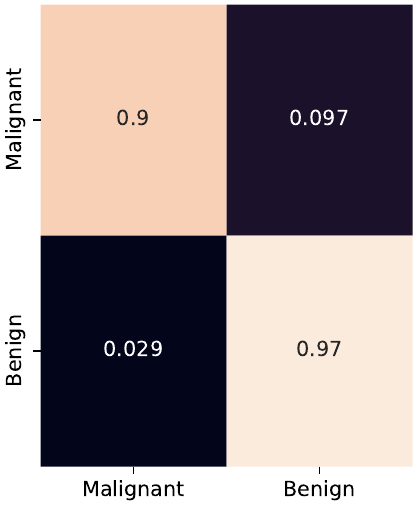}}
		\hfill
		\subfigure[200x]{\includegraphics[width=0.23\textwidth]{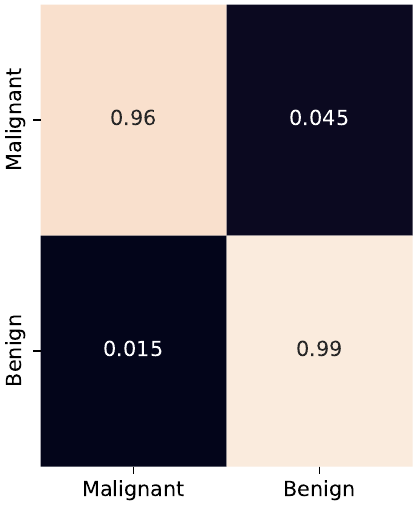}}
		\hfill
		\subfigure[400x]{\includegraphics[width=0.23\textwidth]{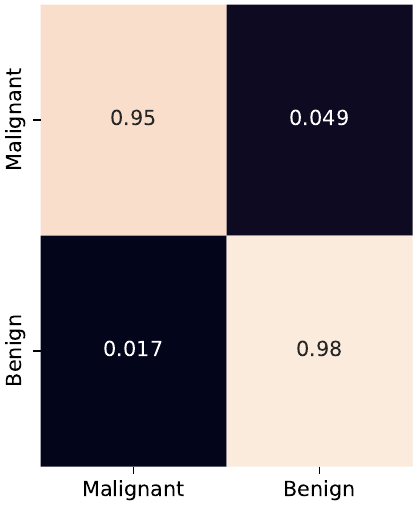}}
		\caption{Confusion matrices for the proposed slim model with G=2 for 40x, 100x, 200x, and 400x zoom factors.}\label{S2CMs}
	\end{figure}

\begin{figure}[h!]
		\centering
		\subfigure[40x]{\includegraphics[width=0.23\textwidth]{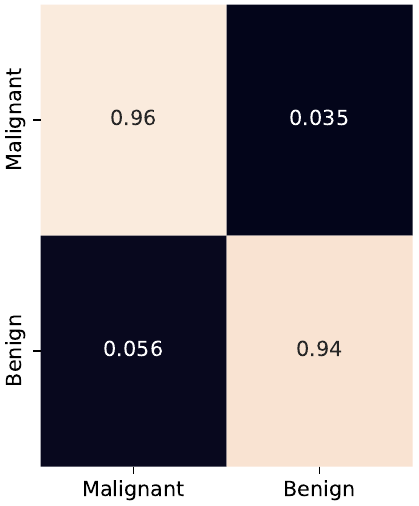}}
		\hfill
		\subfigure[100x]{\includegraphics[width=0.23\textwidth]{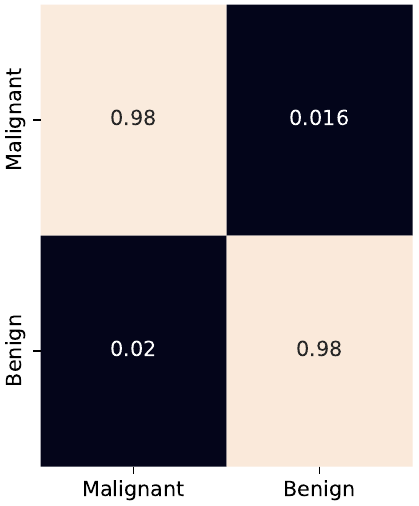}}
		\hfill
		\subfigure[200x]{\includegraphics[width=0.23\textwidth]{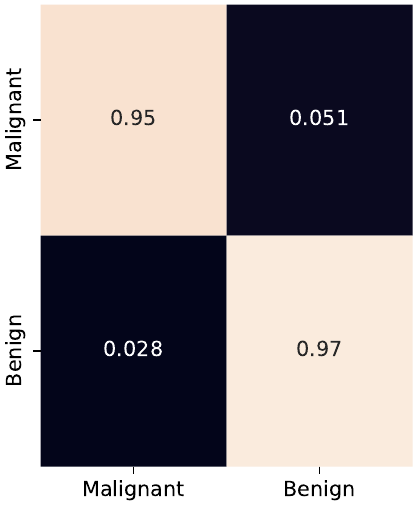}}
		\hfill
		\subfigure[400x]{\includegraphics[width=0.23\textwidth]{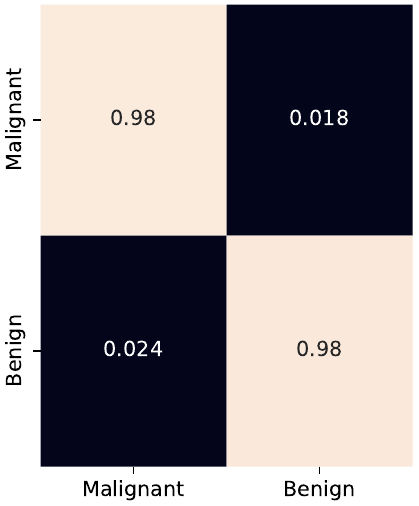}}
		\caption{Confusion matrices for the proposed slim model with G=4 for 40x, 100x, 200x, and 400x zoom factors.}\label{S4CMs}
	\end{figure}

\paragraph{}
The proposed method performance was evaluated by the standard criteria: accuracy, precision, recall, and $F_1$ score, which are described in Equations $4-7$, respectively:
\begin{align}
	\text{Accuracy} &= \frac{TP + TN}{TP+FP+TN+FN}\\ \text{Precision} &= \frac{TP}{TP+FP} \\ \text{Recall} &= \frac{TP }{TP+FN} \\\ F_1 \text{ score} &= \frac{\text{Precision}\times\text{Recall}}{\text{Precision}+\text{Recall}},
\end{align}
Where $TP, TN, FP$, and $FN$ are true positive, true negative, false positive, and false negative rates, respectively. The confusion matrices for the base model, slim model with G$=2$, and slim model with G$=4$ models under all four magnifications are presented in Figures \ref{BCMs}-\ref{S4CMs} respectively.
	
\paragraph{}
We considered popular and state-of-the-art methods to compare our work with, including a 5-layered CNN and the LeNet \cite{LeNet} taken from \cite{Razi}, VGG-19 \cite{VGG}, and ResNet-18 \cite{original_resnet} taken from \cite{DRDA}. We included the best versions among various implementations of these networks reported in the cited papers. Moreover, some of the cutting-edge methods from very recent literature, including \cite{fusion,Spanhol,Spanhol2,Bayramogl,Nahid2,DRDA} were also selected for comparison. \pagebreak \\
\paragraph{}
The classification accuracy of all these methods were compared to the proposed approach (GroupMixer) in Table \ref{Table1}. However, not all papers have reported precision, recall, and $F_1$ score metrics. It can be seen that the proposed method outperforms almost all methods under all magnification factors, except perhaps \cite{Razi, MResNet, ResNet18-TL}, which performed on par with the proposed approach. However, one should also consider the number of parameters for each method. While the compared methods train millions of parameters, the proposed approach optimizes only about $100K$ parameters, which is orders of magnitude smaller than other methods and has the potential for deployment in low-cost embedded devices. Demonstrating state-of-the-art performance in HBC image classification with such a small number of network parameters is a major impact of the proposed approach. Table \ref{table:complexity} shows the number of trainable parameters for each architecture presented in current research. From the results, it can be understood that both the proposed GroupMixer architectures and base model, which is designed based on \cite{Patches}, work very well and outperform lots of previous works in the literature and perform approximately on par with state-of-the-art previous research. Besides, the results are achieved by much fewer parameters than state-of-the-art works, which denotes that our methodology is promising, and probably by searching over hyperparameters, the results can even get better and achieve state-of-the-art performance.
	
\begin{table*}[h!]\centering
	\caption{Summary of Experiments}
	\small\addtolength{\tabcolsep}{-4pt}
	\fontsize{7.5}{9}\selectfont
	%\begin{scriptsize}
	\begin{tabular}{lccccccccccccccccccccc}
	\toprule
	Method  &
			\multicolumn{4}{c}{Accuracy (\%)}
			&&    \multicolumn{4}{c}{Precision (\%)}
			&&   \multicolumn{4}{c}{Recall (\%)}
			&&   \multicolumn{4}{c}{$F_1$ score (\%)}
			\\ \cline{2-5} \cline{7-10} \cline{12-15} \cline{17-20}
			
			&\multicolumn{1}{c}{40X}
			&   \multicolumn{1}{c}{100X}
			&\multicolumn{1}{c}{200X}
			&   \multicolumn{1}{c}{400X}
			&&\multicolumn{1}{c}{40X}
			&   \multicolumn{1}{c}{100X}
			&\multicolumn{1}{c}{200X}
			&   \multicolumn{1}{c}{400X}
			&&\multicolumn{1}{c}{40X}
			&   \multicolumn{1}{c}{100X}
			&\multicolumn{1}{c}{200X}
			&   \multicolumn{1}{c}{400X}
			&&\multicolumn{1}{c}{40X}
			&   \multicolumn{1}{c}{100X}
			&\multicolumn{1}{c}{200X}
			&   \multicolumn{1}{c}{400X}
			&& \\  \hline
			CNN-5  \cite{Razi} & 80.09 & 79.72 & 79.34  & 77.87 && - & - & - & - && - & -& -& - && - & - &- &  - \\
			LeNet \cite{LeNet,Razi} & 80.02 & 79.73 & 79.41  & 78.07 && - & - & - & - && - & -& -& - && - & - &- &  - \\
			Spanhol et al. \cite{Spanhol} & 89.6 & 85.0 & 84.0  & 80.8 && - & - & - & - && - & -& -& - && - & - & - &  - \\
			Spanhol et al. \cite{Spanhol2} & 84.6 & 84.8 & 84.2  & 81.6 && - & - & - & - && - & -& -& - && 88.0 & 88.8 & 88.7 &  86.7 \\
			Bayramoglu et al \cite{Bayramogl} & 83.0 & 83.1 & 84.6  & 82.1 && - & - & - & - && - & -& -& - && - & - & - &  - \\
			%Nahid et al. \cite{Nahid} & 94.40 & 95.93 & {97.19}  & {96.00} && 94.00 & \textbf{98.00} & \textbf{98.00} & {95.00} && 96.00 & 96.36& {98.20}& \textbf{97.79 }&& 95.00 & \textbf{97.00} & \textbf{98.00} &  {96.00} && ?\\
			Nahid et al. \cite{Nahid2} & 90 & 90 & 91  & 90 && 96 & 91 & 93 & 92 && - & -& -& - && 93 & 93 & 93 &  93 \\
			FE-BkCapsNet \cite{fusion} & 92.71 & 94.52 & 94.03  & 93.54 && - & - & - & - && 92.15 & 95.16 & 94.34 &  94.06  && - & - & & - \\
			ResNet-18 \cite{DRDA} & 92.12 & 93.22 & 92.29  & 91.01 && 93.11 & 91.58 & 91.31 & 90.77 && 92.21 & 92.25& 92.19& 91.15 && 92.66 & 91.91 & 91.75 &  90.58 \\
			VGG-19  \cite{DRDA} & 92.11 & 92.00 & 92.00  & 93.45 && 93.21 & 91.15 & 90.96 & 94.00 && 91.65 & 93.44& 91.11& 91.17 && 92.42 & 92.28 &91.03 &  92.56 \\
			{DRDA-Net7 \cite{DRDA}} & 95.72 & 94.41 & 97.43  & 96.84 && 94.00 & {96.00} & 96.00 & 98.10 && 96.90 & 93.20 & 99.00& 95.20 && 95.40 & 94.60 & 97.44 &  96.62 \\
			{Student-40X \cite{Razi}} & 99.31 & 99.30 & 99.23  & 99.14 && 99.27 & 99.17 & 99.01 & 99.00 && 99.30 & 99.17& 99.00& 98.98 && 99.31 & 99.15& 99.02& 98.96 \\
			{Modified ResNet \cite{MResNet}} & 99.54 & 99.11 & 99.52  & 98.74 && 99.18 & 98.61 & 98.92 & 97.89 && 99.37 & 98.86& 99.03& 97.37 && - & - & -& - \\
			{TL-ResNet-18 \cite{ResNet18-TL}} & 99.25 & 99.04 & 99.00  & 98.08 && 99.63 & 98.99 & 98.94 & 98.00 && 96.26 & 99.66& 99.65& 99.19 && 99.44 & 99.33 & 99.29& 98.59 \\
			{TL-GoogLeNet \cite{91}} & 97.89 & 97.64 & 97.56  & 97.97 && - & - & - & - && - &- & -& - && - & - & -& - \\
			{TL-VGG8-patient level \cite{Rahhal}} & 86.20 & 85.90 & 87.2  & 86.3 && - & - & - & - && - &- & -& - && - & - & -& - \\
			\hline
			Presented-Base & 97.65 & 98.92 & 99.21 & 98.01 && 98.91 & 97.88 & 99.63 & 95.90 && 97.65 & 98.92& 99.21& 98.01 && 98.25 & 98.378& 99.42& 96.85 \\
			Proposed-Slim-G2 & 95.85 & 93.66 & 97.01 & 96.71 && 96.21 & 94.58 & 97.37 & 97.10 && 95.85 & 93.66 & 97.01& 96.71 && 96.02 & 94.09 & 97.19 & 96.90 \\
			Proposed-Slim-G4 & 95.42 & 98.16 & 96.05 & 97.92 && 92.91 & 97.27 & 95.61 & 97.05 && 95.42 & 98.16 & 96.05& 97.92 && 94.03 & 97.70 & 95.83 & 97.46 \\
			\bottomrule
			\label{Table1}
		\end{tabular}
		%\end{scriptsize}
	\end{table*}
	
	\begin{table}[h!]\centering
		\caption{Complexity of architectures}\label{table:complexity}
		\begin{tabular}{|c|c|c|c|}
			\hline
			Model  & Base & Slim (G$=2$) & Slim (G$=4$) \\
			\hline
			Number of Parameters & 102018 & 77442 & 65154 \\
			\hline
			\iffalse
			\hline
			f-measure & 98 & 99 & 98 \\
			\hline
			\fi
		\end{tabular}
	\end{table}
	
\section{Conclusions}\label{sec concl}
In the current research, we utilized some of the advancements of recent works and integrated a simple CNN architecture with patch embedding operation to achieve transformer-level performance with a low-complexity CNN architecture while avoiding quadratic computational complexity of transformers and taking into account the disability to train transformer-based architectures due to lack of large medical image datasets, especially lack of large HBC datasets. Although the presented architecture has much fewer parameters than competitors, the presented architecture achieved state-of-the-art performance and outperformed lots of recent works in the literature. We took a step forward and proposed a slim architecture using Group Convolution and Channel Shuffling ideas and reduced the number of trainable parameters even more with a slight performance reduction. Achieving such great performance with extremely small number of parameters is remarkable, and we hope to attract other researchers attention to developing high performance light deep neural network architectures without a decline in performance, which is a necessity due to the lack of large medical datasets.
	
\end{document}